# Simultaneous Amplitude and Phase Spectroscopy using Two-Photon Interference

Kyle M. Jordan, Yingwen Zhang, Frédéric Bouchard, Duncan England, Philip J. Bustard, Benjamin J. Sussman(*), Jeff S. Lundeen(*), Andrew H. Proppe(*)

**Abstract**

Quantum spectroscopy seeks to probe chemical systems using nonclassical light, which has properties that are qualitatively and quantitatively different than conventional light sources. One promising technique uses intensity-correlated twin beams of light to reduce the noise sources inherent to absorption spectroscopy. However, measurements of the phase shift imparted by the chemical sample, which provides complementary information to the absorption, continue to be a challenge. Here, we propose and demonstrate a scheme using entangled photon pairs that can simultaneously measure both the absorption and phase shift of a sample with extremely low optical intensities and with relatively fast few-minute acquisition times. This method combines the previous use of intensity correlations with a broadband quantum interferometer utilizing two-photon interference to measure the complete linear optical response of the sample. Our work shows that precise measurements of absorption can also be made phase-sensitive using suitable choices of probe beam and detection scheme. This enables a new class of quantum spectroscopy schemes which measure absorption and phase with a single probe. Our technique is relevant to the characterization of a wide array of chemical and biological samples, such as quantum dots and organic fluorophores, and may be useful for spectroscopic measurements that are otherwise constrained in intensity.

**Introduction**

Our knowledge of the electronic structure of molecules and materials comes largely from careful measurements of absorption and emission spectra, from which the various quantum energy levels of a sample can be determined. Commonly, only the amplitude of the outgoing light is measured, since this is easily accomplished using a spectrometer; various peaks and linewidths in the measured signal can be used to infer electronic properties of interest. However, the optical phase shifts imparted by the sample contain complementary information, which is particularly relevant in the study of dynamical processes. For instance, measurement of these phase shifts by spectral interferometry has been used to disentangle thermal and electronic contributions to resonance linewidths, which cannot be determined from absorption alone[1]. Time-resolved (i.e. pump-probe) absorption spectroscopy, which is fundamentally a type of nonlinear interferometry, provides insight into coherent dynamics occurring on ultrafast timescales, and is useful to study the dynamics of vision[2] as well as materials used in light-emitting and light-harvesting technologies[3,4]. Although time-resolved spectroscopy is commonly used to measure phase shifts, the use of intense laser pulses can lead to unwanted changes in the sample of interest. For example, many organic molecules and single emitters experience irreversible photodamage after sufficient exposure to light[5–7]; other effects, such as photocatalysis, may also hinder the measurement. In these circumstances, phase-sensitive spectroscopic methods which operate with a low photon flux become important.

In situations where the number of incident photons is constrained, it becomes essential to consider the amount of spectroscopic information which is gained per probe photon used. For the case of absorption spectroscopy, the sample's absorption at a given wavelength is determined by the ratio of the intensity incident on a sample to the detected intensity. Determining the absorption of a sample therefore relies on precise measurements of the intensity of the probe beam. In the absence of any technical noise sources, absorption spectroscopy is limited in precision only by the intensity fluctuations arising from the quantum nature of light[8,9]. Of all classical light sources, a laser operated far above threshold results in the smallest intensity fluctuations for any fixed mean intensity[10]. The intensity signal-to-noise ratio of such light is $\sqrt{N}$ when the beam contains an average of $N$ photons, which places an upper bound on the precision of any intensity measurement known as the shot noise limit (SNL). For photosensitive samples, there is an upper limit to the number of photons which may be used to probe the sample, which in turn limits the precision of classical absorption spectrometry via the SNL. On the other hand, quantum light, consisting of twin beams which have correlated intensities, can surpass this limit. Typically, one beam (the probe) interacts with the sample, while the other (the herald) is directly measured with a photodetector[11–16]; due to the strong intensity correlations between the beams, measurements of the herald's intensity can (ideally) predict with certainty the intensity of the probe beam, resulting in more precise absorption measurements. Furthermore, rigorous calculations[17,18] have shown that this technique, combined with sufficiently high-efficiency detectors, results in the highest precision-per-photon of any absorption measurement allowed by quantum mechanics. Recent experiments using this technique have demonstrated improved precisions up to 40% higher than the SNL[12,14]. At the same time, as in traditional absorption spectroscopy, these heralded measurements are entirely insensitive to the phase shift imparted by the sample. This paper presents a heralded absorption technique that enables direct measurements of phase while still preserving the strong intensity correlations that outperform the sensitivity of traditional absorption spectroscopy.

The linear optical response of a sample is fully described by the absorption and phase spectrum it imparts on a probe beam. When a spectroscopic sample is exposed to a monochromatic optical field with amplitude $E_0$ and frequency $\omega$, the transmitted field generally has an amplitude $E_0 e^{-A(\omega)/2 + i\phi(\omega)}$, where $A(\omega)$ and $\phi(\omega)$ define the absorption spectrum (quantified by the absorbance) and phase spectrum, respectively. Common methods of phase-sensitive spectroscopy that use classical light include spectroscopic ellipsometry[19], which is useful for samples with optically smooth surfaces, and white-light interferometry[20–22]. Approximate methods using the Kramers-Kronig relations[23] and iterative matrix inversion[24] have been applied to estimate the complex dielectric function of colloidal quantum dots from absorption and photoluminescence measurements. These methods rely on the fact that the absorption and phase response of a sample are interdependent quantities; knowledge of the linear absorption spectrum of a sample at all frequencies can be used to make predictions about its phase response, and vice versa[25]. While this works particularly well for resonances, there is generally not a one-to-one relationship between absorption and phase response, so that using absorption to predict phase can be unreliable[26]. In the case of light-sensitive samples, the phase response gains a new significance: since both the absorption and phase response contain information about resonance features, a joint measurement of both quantities can in principle improve the precision in estimates of resonance parameters when compared to measurements of absorption or phase alone[27]. To maximize the information gained from a given number of probe photons, it becomes important to not only optimize the sensitivity of the absorption measurement but also to extract some phase information, so long as one does not come at the cost of the other.

Measurement of phase shifts using correlated photons is possible by two means. First, the sample can be placed in an interferometer (*e.g.*, a Mach-Zehnder interferometer), through which the probe photon travels. This effectively replicates the classical white-light interferometer, with a herald photon in a separate beam path[28]. Such a method suffers from the same experimental challenges as white-light interferometry, mainly the requirement for precise phase stabilization throughout the measurement process. A second method, which we use, relies on two-photon Hong-Ou-Mandel (HOM) interference between the probe and the herald photons, which occurs when sending one photon to each of the two input ports of a beam splitter[29]. This interference occurs only if the probe and herald photons overlap in all degrees of freedom: time, frequency, spatial mode and polarization. In this case, the physical outcome is the same if both photons transmit through the beam splitter or if both photons reflect at the beam splitter, leading to quantum interference. This interference appears as Ramsey-like interference fringes in the rate at which both photons exit opposite beam splitter port[30]. Due to energy conservation, complementary fringes are seen in the rate at which both photons exit through the same beam splitter port (*i.e.*, photon bunching). In the simplest scenario, HOM interference occurs between two unentangled and indistinguishable photons, leading to a complete suppression of coincidence detections, in which detectors at the two output ports both click. When frequency entangled photons are used instead, HOM interference either enhances the probability of a coincidence detection or suppresses it, depending on the relative phases encoded in the two-photon quantum state[31]. Specifically, if one photon has frequency $\omega_1$ and the other $\omega_2$, then the rate at which pairs bunch and antibunch is proportional to

$$|S(\omega_1, \omega_2) \pm S(\omega_2, \omega_1)|^2,$$

where the terms add for bunching, and subtract for antibunching. Here,

$$S(\omega_1, \omega_2) = e^{-A_a(\omega_1)/2 - A_b(\omega_2)/2 + i\phi_a(\omega_1) + i\phi_b(\omega_2)}$$

is the probability amplitude of a photon pair consisting of frequencies $\omega_1$ and $\omega_2$, where the $\omega_1$ photon experiences an absorbance $A_a$ and phase shift $\phi_a$ and the $\omega_2$ photon experiences an absorbance $A_b$ and phase shift $\phi_b$. This HOM method has the advantage of being less sensitive to relative changes in the path lengths of the two interferometer arms compared to a Michelson or Mach-Zehnder interferometer since fringes depend on the phase difference between two frequencies rather than the absolute phase. Stable interference fringes can therefore be observed between photon pairs with wavelengths separated by more than 100 nm without the use of active phase stabilization[31,32].

This article introduces a modification to previous absorption-only quantum spectroscopy schemes[12–14], enabling measurements not only of the absorption (quantified by the absorbance $A(\omega)$) of a sample but also its phase shift $\phi(\omega)$ using a single probe beam. It achieves this by using a probe beam that is entangled in both intensity and frequency with a herald beam. The absorption spectrum is measured using intensity and frequency correlations between the beams. As illustrated by Figure 1(a), absorption of a photon at a particular frequency (red, in the figure) results in a single detector click at the corresponding frequency (blue) of the herald photon. The phase spectrum is measured using HOM interference[29,30,32] between the two beams, in which the exit port of each photon from the beam splitter as well as its frequency is recorded (Figure 1(a)). In this way, absorption and phase are measured independently, with no need to infer one from the other. We demonstrate this scheme by measuring the infrared absorption and phase spectra of dye molecules (silicon 2,3-naphthalocyanine dichloride (SiNC)) with a few picojoules of quantum-correlated probe light and few-minute total exposure times. By probing both absorption and phase, our method allows for model-free determination of the complete linear optical behavior of a sample. The low photon flux of the probe beam makes this precision spectroscopy technique suited to measurement of light-sensitive samples such as organic molecules and single emitters.

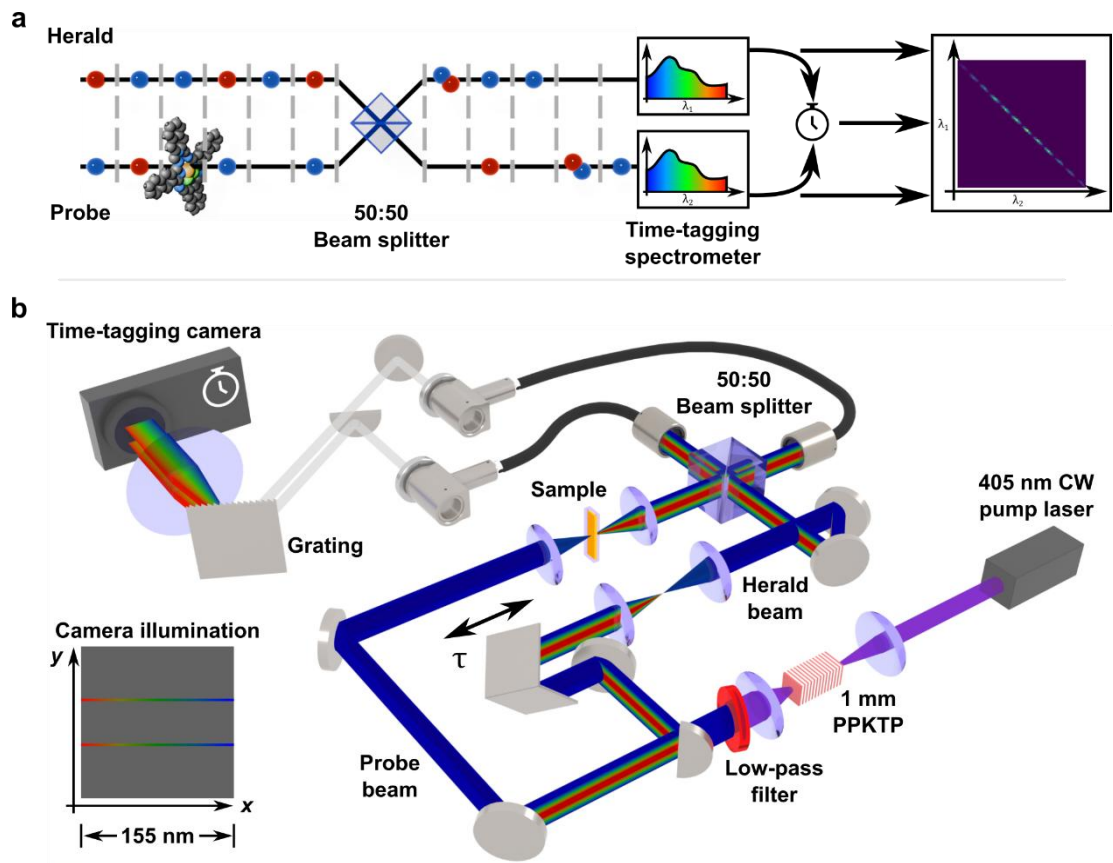

**Fig. 1**: **a**, Schematic diagram of the measurement. Photon pairs are prepared and separated into two intensity- and frequency-correlated beams. The frequency correlations ensure that knowledge of one photon's frequency (say, red) also determines the other photon's frequency (say, blue). One beam interacts with a spectroscopic sample, after which both beams are mixed

at a beam splitter; each output port is sent to a time-tagging spectrometer. Detected photons are categorized depending on whether they arrive individually or as part of a time-correlated pair. Pair detections are further categorized based on which beam splitter output port each photon emerged from. Due to the frequency correlations, an unpaired click at one frequency (here, blue) indicates that the sample absorbed a photon at the corresponding frequency (red). The sample's absorption spectrum is determined from the ratio of individual to paired detections; the optical phase spectrum of the sample is determined by the fraction of pairs exiting the beam splitter with a single photon at each output port, which exhibits a phase-sensitive interference pattern. **b**, Experimental spectroscopy scheme. Quantum-correlated photon pairs are generated through SPDC, with one photon being used to probe a chemical sample and the other photon being used as a herald. Hong-Ou-Mandel interference between the two photons occurs at a beam splitter, after which both photons are sent to a two-photon spectrometer, which resolves the joint spectrum of each photon pair.

## Results

Our spectroscopy scheme consists of three main stages (details are provided in the Methods section). First, spontaneous parametric down-conversion (SPDC) is used to prepare two broadband, highly entangled beams of light, which we label the probe and herald beams. Due to the relatively low pump power used to drive the SPDC, the down-converted light consists of a stream of frequency-entangled photon pairs, with one photon in each of the two beams. Second, we expose a sample to the probe beam and subsequently interfere both beams at a beam splitter. Finally, we perform a spectrally resolved measurement of the two beam intensities using a two-photon spectrometer.

*Performing a spectroscopic measurement*

We introduce only the probe beam to the spectroscopic sample. A controlled delay $\tau$ applied to the herald beam provides a variable phase shift $\omega\tau$ at each frequency. The two beams are then mixed at a 50:50 beam splitter so as to induce two-photon HOM interference, which provides the phase-sensitive signal to the spectrometer.

In frequency-resolved HOM, there are two interfering paths that lead to detection of two photons after the beam splitter with frequencies $\omega_1$ and $\omega_2$: either the probe photon had frequency $\omega_1$ and the herald had $\omega_2$, or vice versa. The phases of these interfering paths are respectively $\phi(\omega_1) - \omega_2\tau$ and $\phi(\omega_2) - \omega_1\tau$. The bunching and antibunching count rates exhibit interference fringes that depend on the relative phase of the paths and that have a period of $2\pi/|\omega_1 - \omega_2|$ when the delay $\tau$ is varied. Explicitly, the coincidence rate for photon pairs at frequencies $\omega_1$ and $\omega_2$ is

$$R_{coinc} \propto [T_s(\omega_1)T_r(\omega_2) + T_r(\omega_1)T_s(\omega_2)](1 \pm V \cos[\phi(\omega_1) - \phi(\omega_2) - (\omega_2 - \omega_1)\tau]),$$

where $T_s$ and $T_r$ denotes the transmission of the sample and reference arms, respectively, and $V$ is the interference visibility, which depends on the transmission of each arm and on the spatial mode overlap of the two photons. The final term in brackets is positive when considering bunching pairs and negative when considering antibunching pairs (see Supplementary Note 1 for details). When considered a function of $\tau$, the sample-dependent phase $\phi(\omega_1) - \phi(\omega_2)$ acts as a constant shift in the position of the interference fringes. As a result, the ratio of bunched pairs to antibunched pairs shows sinusoidal oscillations as a function of $\tau$ from which we determine the phases imparted by the sample.

The measured interferogram shows a symmetry under the interchange $\omega_1 \leftrightarrow \omega_2$, which is a consequence of the indistinguishability of probe and herald photons after the beamsplitter. The measured phase shift for a pair of frequencies $\omega_1, \omega_2$ contains contributions from the sample response at both $\omega_1$ and $\omega_2$, which causes some ambiguity in the reconstruction of asymmetric phase responses. The SiNC sample used in our measurement is expected to have a flat phase response in one half of the available bandwidth, so that the phase shifts obtained from the interferogram can be attributed entirely to the other half of the measured bandwidth. Methods which allow for reconstruction of more complicated asymmetric sample responses from the measured interferograms are considered in the Discussion.

Importantly, information about the absorption spectrum is provided at the same time by the number of transmitted photons as compared to the number of absorbed photons. After the beam splitter, the probe and reference photons can no longer be distinguished due to the loss of path information; nevertheless, detections can be separated into pair ("coincidence") detection and unpaired ("single") detections, the distribution of which encode information about the absorption. As in previous quantum absorption spectroscopy methods[12,14], a coincidence detection necessarily means that the sample transmitted the probe photon; meanwhile, single detections result either from the sample's absorption or from optical losses. The sample absorption can then be determined via the ratio $c^{(s)}/c^{(r)}$, where $c^{(s)}$ and $c^{(r)}$ denote the measured coincidence-to-singles ratio in the case of the sample of interest $(s)$ and the reference sample $(r)$, respectively. Due to the presence of the beamsplitter, this ratio is not directly proportional to the sample transmittance $T_s$, and certain calibration measurements must be made to correctly estimate $T_s$; this procedure is described in Supplementary Note 1. In this way, the spectrally resolved HOM interference pattern, together with the spectrum of single photons, yields separate signals for the absorption and phase of the sample.

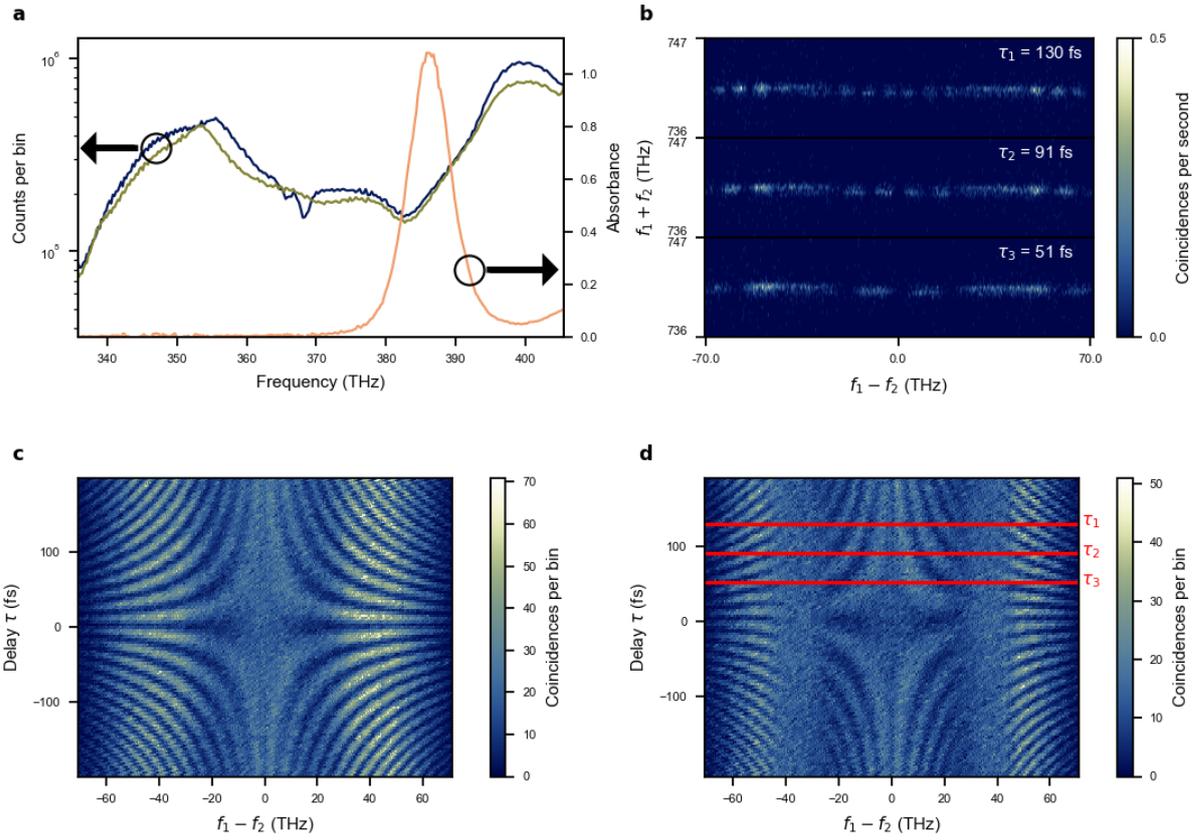

**Fig. 2:** Probe spectrum, Hong-Ou-Mandel interference, and frequency correlations. **a**, Spectrum of individual photons collected into the two fibers (blue and green). These distributions include both paired and unpaired photons. The absorbance spectrum, in absorbance units, of the SiNC dye is shown in orange, as measured by a conventional spectrometer. **b**, Joint spectrum of the photon pairs that antibunched at the beam splitter. The relative delay between probe and herald photons is displayed for each spectrum. **c, d**, Spectrally resolved Hong-Ou-Mandel interference pattern as a function of frequency difference and relative delay. This interferogram is shown for both the reference (**c**) and for the dye (**d**) samples. Red lines in **d** indicate the three delays depicted in **b**.

*Determining the absolute spectral response of a sample*

To perform spectroscopy of a single sample, the relative delay $\tau$ is stepped by 1.2 μm (4 fs) across a total range of 120 μm (400 fs). At each delay, we record a three second exposure, resulting in a total exposure time of five minutes per measured sample. As Figure 2(b) shows, each exposure records a pattern of HOM interference fringes which vary as a function of the measured frequency difference $f_1 - f_2$. The interference fringes at a fixed pair of frequencies oscillates as a function of the delay $\tau$ with a period of $1/|f_1 - f_2|$. Figures 2(c) and 2(d) show respectively the joint spectra of antibunched light for the reference and SiNC samples as a function of both frequency difference and delay, so that each row of the image corresponds to the HOM fringes at a fixed delay; the corresponding images for bunched light shows inverted fringes. Phase shifts from the sample displace the positions of the HOM fringes relative to the reference measurement at each vertical slice of the frequency-delay plots (Figures 2(c) and 2(d)), while absorption reduces both the joint intensity and the fringe visibility. When no absorbing sample is present, each three second exposure records an average of approximately 216,000 single detections and 976 coincidence detections, which in the current experiment is limited solely by saturation of the sensor and the collection efficiency. This corresponds to a measured power of 18 femtowatts, or approximately 350 femtowatts of power produced by the source in the pair of spatial modes we use for measurements. The current source configuration exposes the sample to additional power due to the SPDC source's many spatial modes, only one pair of which are detected. In measurements with a photosensitive sample, this unwanted exposure can be easily prevented by using a fiber-coupled source of photon pairs containing only a single pair of correlated spatial modes. Exposure times can be improved by spreading the beam over multiple pixel rows using a pair of cylindrical lenses, which then enables the use of a brighter source of photon pairs. The intensity of the probe beam is ultimately limited by the requirement that only a single pair of photons is generated within the coherence time of the probe beam.

To demonstrate the new scheme, we perform a differential measurement of a 0.15 mM solution of SiNC dye molecules in a toluene solution, relative to a sample containing only toluene. We repeat the measurement procedure detailed above a total of ten times for each of the two samples and use the standard deviation of the resulting transmission and phase spectra to estimate the measurement precision. The differential transmission and phase spectra of the SiNC dye obtained using the coincidence counts are shown in Figures 3(a) and 3(b), respectively. We compare these spectra to the transmission spectrum of the same dye measured with a conventional spectrometer (Duetta, Horiba) and a phase spectrum calculated using the Kramers-Kronig relations. We also show the transmittance obtained from the ratio of singles counts with and without the sample, effectively mimicking a classical absorption spectrometer using our quantum light. The systematic errors in both spectra are attributed primarily to deflection of the SPDC light due to wedged cuvette glass; this deflection has a similar effect on the coincidence counts and on the singles counts, so that the transmission curve obtained from the coincidence-to-singles ratio and the curve obtained from the singles spectrum both show similar errors (see Supplementary Note 2 for further discussion). Practical use of this technique requires careful consideration so as to minimize this effect. The quantum spectroscopy technique is able to recover the qualitative features of both spectra and has quantitative agreement with the peak absorption as well as the peak phase shift.

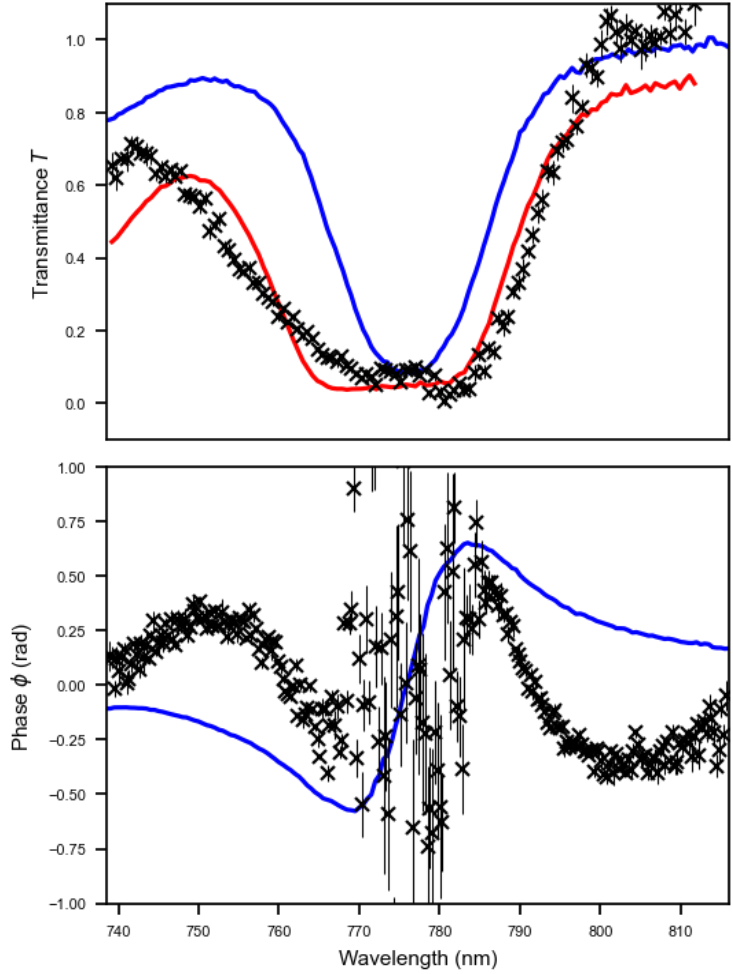

**Fig. 3:** Estimated transmittance $T$ and phase $\phi$ profiles for SiNC dye. The estimated transmission and phase profiles obtained from the coincidence-to-singles ratio and the HOM interferogram, respectively, are shown in black. A red curve shows a transmission estimate obtained using only the spectrum of single counts. The blue curves show the transmission as measured by a conventional spectrometer, in absorbance units, and a corresponding phase profile determined by the Kramers-Kronig relations. Error bars show the standard error calculated from ten repeated measurements.

**Discussion**

The spectroscopy method discussed herein introduces the concept of simultaneous absorption- and phase-resolving measurements to the field of quantum spectroscopy. The HOM scheme can be seen as the merger of two existing techniques: first, the use of heralded photon pairs for absorption spectroscopy, and second, the use of HOM interference to probe phase shifts. The two methods have been demonstrated separately in previous experiments[12,14,32–36]. This work demonstrates for the first time that both quantum techniques can be combined into a single measurement to recover both absorption and phase information.

Two-photon HOM interference has long found applications in non-spectrally resolved measurement of short time delays[29,31,37] due to both its phase stability[38] and to the large signal-to-noise ratio possible using coincidence detection of photon pairs[37]. As an extension of this, recent work has sought to use (non-spectrally resolved) HOM interference to infer the presence of molecular absorption due to the phase shifts imparted on the photons[33]. It has also been suggested that the use of frequency-entangled pairs in HOM interference can improve the resolution of time-dependent molecular processes[33]. Our results make it clear that spectrally resolved HOM interference, which also yields the traditional HOM signal upon suitable integration of the results, provides additional information which has not been exploited in previous techniques. In particular, most HOM experiments to date focus only on pairs of photons which both arrive at the detector, discarding any information contained in the ratio of coincidence detections to single detections. Our results show that information about the sample's absorption may be extracted from these extra counts without significant changes to the standard HOM interferometer. This analysis, or variations thereof, can be easily applied to many existing HOM-based methods to enable sensing of absorption and loss while also measuring other quantities.

Unlike traditional spectroscopy, the phase spectrum measured via HOM interference is symmetric due to the indistinguishability of the probe and herald photons after the beam splitter. The current experiment reconstructs the spectrum only on one side of the center frequency $\omega_1 = \omega_2 = \omega_{pump}/2$; in more complicated samples, the phase response may extend over the entire measured bandwidth, so that general methods of resolving this ambiguity become necessary. We propose two possible solutions to this problem. The absorption spectrum obtained from the singles counts is generally not symmetric, so that a joint fit of both the singles and coincidence spectra to a model of $A(\omega)$ and $\phi(\omega)$ can be used to detect asymmetric features in the phase spectrum. If a model of the sample is not available, then multiple measurements using different pump wavelengths can be performed. Each measurement will yield a phase response which has been symmetrized about a different center frequency, so that the collection of all measured responses can be used to reconstruct an asymmetric sample response. This HOM spectroscopy technique is therefore not restricted to simple molecules such as the SiNC dye we use, and can be applied to samples with arbitrary spectral responses.

For practical spectroscopy, three features are desirable: 1) a broad bandwidth, with fine spectral resolution; 2) short exposure times; and 3) a large signal-to-noise ratio (SNR). The photon pair source we use has a bandwidth in excess of 150 nm, which is primarily limited by the bandwidth of our optics and by the 900 nm cutoff wavelength of the detector. With a suitable choice of detector and optics, bandwidths extending further into both the infrared and visible frequencies are possible. In particular, the bandwidth of light produced by a PPKTP crystal is primarily limited by the crystal's thickness and poling structure; we use a 1 mm thick crystal with uniform poling, but crystals with chirped poling can produce octave-spanning bandwidths[39]. As mentioned above, our resolution of 0.6 nm is pixel-limited, so that improvements to resolution must involve a higher-resolution camera[40,41], illumination of additional pixels, or suitable processing of the data to achieve sub-pixel resolution[42]. The exposure time of our measurement is currently limited due to detector saturation, which can be improved by illuminating a larger number of pixels.

From theoretical analysis of other heralded absorption measurements, it is expected that low system losses are critically important in demonstrating a quantum enhancement in precision[12]. Since phase information is obtained only from transmitted photons, low losses are also beneficial for phase measurement. Our experiment is limited in this sense by a low system efficiency of about 0.5%, averaged across all spectral bins; this is primarily due to the quantum efficiency of our detector (7%)[43] and due to losses involved in coupling the photon pairs into single-mode fiber. This low system causes many more single clicks to be recorded than coincidence clicks, so that estimates of absorption and phase which rely on the measured coincidence rates have large statistical uncertainties in our measurement. Improvements are possible by using a detector with an improved quantum efficiency[40,44,45] and, also, by reducing system losses. In particular, time-of-flight spectrometers which use superconducting nanowire detectors have quantum efficiencies which are set mainly by loss in the dispersive fiber; depending on the choice of fiber, the efficiency of such a spectrometer can be almost ten times that of our time-tagging camera. System losses in our experiment are due mainly to the difficulty of coupling broadband correlated photon pairs into single mode fiber. Suitable engineering of the source and collection optics has resulted in narrowband fiber-coupled SPDC sources with near-perfect efficiency[46]. While the broadband sources useful in spectroscopy are more challenging to design, we expect that broadband photons similar to the ones used here can be coupled into fiber with greater than 70% efficiency. Based on similar heralded absorption measurements[12], the quantum enhancement in absorption precision is expected to grow as the overall loss decreases, with significant precision enhancement occurring for system efficiencies above 50%.

Quantum-enhanced spectroscopy is important whenever the total amount of probe light is limited. Despite many demonstrations of precision improvements for measurements of either phase or absorption, research into simultaneous measurement of multiple physical quantities is still in its infancy. Our work demonstrates the versatility of quantum interferometry for this task, as it can be used to encode separate physical quantities into independent properties of the probe light (here, the rates of heralded detections and the bunching statistics of paired photons). The use of entangled quantum probes can be applied more broadly to other interferometers as well, bringing highly informative and highly sensitive probes to the frontiers of low-intensity measurements.

**Methods**

*Producing entangled photon pairs*

Spontaneous parametric down-conversion (SPDC) is the process by which a single pump photon at frequency $\omega_{pump}$ is split into a pair of photons (probe and herald) with frequencies $\omega_{probe}$ and $\omega_{herald}$ inside a $\chi^{(2)}$-nonlinear medium. We use a 1 mm-long type-0 periodically poled potassium titanyl phosphate (PPKTP) crystal pumped with a 405 nm continuous-wave laser to produce photon pairs with identical polarizations (Figure 1(b)). Due to energy conservation, the three photons must have frequencies related by $\omega_{pump} = \omega_{probe} + \omega_{herald}$, so that the probe and herald photons have strongly correlated frequencies[47,48]. Energy conservation also ensures that the two photons are created within a short time interval[49]; for broadband photons, this can be as small as a few femtoseconds[50]. Since only $\omega_{pump}$ is fixed by the experimental apparatus, down-converted light may be extremely broadband; for instance, octave-spanning sources have been demonstrated[39]. For our choice of crystal and geometry the two down-converted photons have identical spectra centered at 810 nm and measured bandwidths of 155 nm, which is primarily limited by the bandwidth of our optics and detector. Conservation of momentum requires that the down-converted photons are emitted at equal but opposite angles about the pump direction[47,48]. We therefore use a D-shaped mirror to split the cone of light produced by SPDC into two beams[32]. Due to the momentum anticorrelations, each of the two beams contains only a single photon from each pair, so that the intensities and frequencies of the beams are strongly correlated. These same anticorrelations require that we invert one of the beams along both transverse directions (this is accomplished using a retroreflector) to ensure proper spatial overlap between the two photons at the beamsplitter and allow the photons to interfere.

*Measuring joint spectra and two-photon interference*

Both the HOM interference pattern used for phase measurement and the comparison of coincidence and single counts used for absorption measurement require a method of identifying those detection events which correspond to paired photons. This functionality is provided by a recently developed type of photodetector consisting of an image intensifier along with a camera sensor that tags the time-of-arrival for each detection event at each pixel[51,52]. Due to the tight timing correlations resulting from energy conservation, pairs of photons will always arrive at the sensor within a short time interval, which is mainly limited by the time resolution of the detector (in our case, a resolution of 7 ns[43]; this is in contrast to traditional sCMOS cameras, which are limited to thousands of frames per second, or resolutions of about 100 µs). These timing correlations are exploited in our analysis to identify pairs of detector clicks originating from a single emitted photon pair. We also filter the detected pairs according to their frequency correlations to further improve the SNR. To improve the spatial indistinguishability of the two detected photons and hence increase the HOM interference visibility, we perform spatial filtering of the beams using single mode fibers before sending the photons to the spectrometer. A grating and lens are then used to disperse the light from the two fibers onto separate rows of the camera, forming a single photon-sensitive spectrometer for each of the two fibers as was done in the experiment by Zhang *et al*[32] (see Figure 1(b)). The camera used here (Phoebe TPX3CAM, Amsterdam Scientific) has a 256-by-256-pixel sensor, of which we illuminate two individual rows. Our spectrometer has a pixel-limited resolution of 0.6 nm; however, other experiments have demonstrated that sub-pixel resolution is possible with this camera by employing a statistical analysis of the distribution of detection events[42], which could be used to further enhance the spectral resolution.

*Post-processing*

Due to the large volume of data collected in a single measurement (pixel coordinates and time-of-arrival of each photon), we rely on post-processing to retrieve the optical spectra from the raw data. We discriminate between coincidence and single detections by searching for events arriving within a short time interval (25 ns) of each other. At this stage we discard any pairs which show a large violation of the energy conservation condition $\omega_{pump} = \omega_{probe} - \omega_{herald}$, and which therefore cannot belong to the same down-converted pair. Such events occur when photons from two distinct pairs arrive within a single 25 ns window and, also, when a dark count occurs close in time to a real photon detection. Finally, we use the position of each event within one of the two illuminated rows of the camera to determine the wavelength of the photon and the beam splitter port from which it emerged.

In this way we obtain two distinct types of spectra. First, a 1D spectrum of single detection events. Second, two different 2D joint spectra which are provided by the combined frequencies of a coincidence detection event: one joint spectrum contains the number of detections at each pair of frequencies for photons which both emerge from the same fiber ("bunched" pairs exiting the same beam splitter port), and a second joint spectrum contains the same information for photons that emerge from different fibers ("antibunched" pairs exiting opposite ports). Example 1D spectra are shown in Figure 2(a), and example joint spectra corresponding to antibunched detections are shown in Figure 2(b) for three values of the relative delay $\tau$. From the joint spectra, one sees the strong frequency correlations between paired photons, so that all joint detection events occur along a single stripe of the 2D plane. All joint spectra possess a mirror symmetry which is due to the indistinguishability of photon pairs with frequencies ($\omega_1$, $\omega_2$) to those with frequencies ($\omega_2$, $\omega_1$).

**Acknowledgements**

This work was funded by National Research Council Canada's Internet of Things: Quantum Sensors Challenge program (QSP-112-1), the National Sciences and Engineering Research Council (NSERC), the Canada Research Chairs program, and a Canada First Research Excellence Fund award on Transformative Quantum Technologies. The authors thank Michał Lipka and Michał Parniak for helpful discussions and for providing code used for simulations. The authors acknowledge the use of 3D models provided by Thorlabs, Inc. in diagrams.

## 1. Theoretical model, calibration procedure, and estimators

**Modelling the optical losses**

Our calibration procedure serves to correct for additional losses in the optical system, which are primarily due to finite transmission in the interferometer arms, fiber coupling and detector losses. To justify this calibration procedure, we first present a simple model of the effect of losses on the measured photon count statistics.

Our model separates losses before the beamsplitter, occurring in the interferometer arms, from losses after the beamsplitter, due to fiber coupling and detector loss. We denote the transmission in the signal and reference arms by $T_i$ and $R_i$ respectively, where the index $i$ is the spectral bin. The signal arm also imparts a phase shift $\phi_i$. The net transmission after the beamsplitter is $\eta_{Ai}$, $\eta_{Bi}$, where $A$ and $B$ mark the two spatial modes after the beamsplitter. The optical state produced by the crystal is assumed to have the form

$$|\Psi_0\rangle = \sqrt{1-g}|0\rangle + \sqrt{g} \sum_{i,j=1}^{N} \Psi_{ij}\, a_i^\dagger b_j^\dagger |0\rangle, \tag{1}$$

where $a_j^\dagger$, $b_j^\dagger$ are creation operators for spectral bin $j$ in each of the two interferometer arms, $g$ is the probability of a pair emission, and $\Psi_{ij}$ is a spectral amplitude normalized as $\sum_{ij}|\Psi_{ij}|^2 = 1$. Losses in the two arms, including the absorption and phase shift due to the sample, are modelled by the transformations $a_i \mapsto \sqrt{T_i} e^{i\phi_i} a_i + \sqrt{1-T_i}\, a'_i$ and $b_i \mapsto \sqrt{R_i}\, b_i + \sqrt{1-R_i}\, b'_i$ for environment modes $a'_i$ and $b'_i$. The 50:50 beamsplitter performs the transformations $a_i \mapsto \frac{1}{\sqrt{2}} c_i + \frac{1}{\sqrt{2}} d_i$, $b_i \mapsto \frac{1}{\sqrt{2}} c_i - \frac{1}{\sqrt{2}} d_i$, where the new modes $c_i$ and $d_i$ describe the field after the beamsplitter. After performing these transformations and taking a partial trace over all environment modes $a'_i$, $b'_i$, the mixed state of the photons is found to be

$$\rho = \left[1 - g + g\sum_{ij}|\Psi_{ij}|^2(1-T_i)(1-R_j)\right]|0\rangle\langle 0|$$

$$+ \frac{g}{2}\sum_i (1-T_i)\left[\sum_j \Psi_{ij}\sqrt{R_j}(c_j^\dagger - d_j^\dagger)|0\rangle\right]\left[\sum_j \Psi_{ij}^*\sqrt{R_j}\langle 0|(c_j - d_j)\right]$$

$$+ \frac{g}{2}\sum_j (1-R_j)\left[\sum_i \Psi_{ij}\sqrt{T_i}e^{-i\phi_i}(c_i^\dagger + d_i^\dagger)|0\rangle\right]\left[\sum_i \Psi_{ij}^*\sqrt{T_i}e^{i\phi_i}\langle 0|(c_i + d_i)\right]$$

$$+ \frac{g}{4}\left[\sum_{ij} \Psi_{ij}\sqrt{T_iR_j}e^{-i\phi_i}(c_i^\dagger + d_i^\dagger)(c_j^\dagger - d_j^\dagger)|0\rangle\right]\left[\sum_{ij} \Psi_{ij}^*\sqrt{T_iR_j}e^{i\phi_i}\langle 0|(c_i + d_i)(c_j - a_j)\right]. \quad (2)$$

The first term describes the loss of both photons; the second and third terms describe the loss of a single photon; and the final term describes transmission of both photons. The projectors describing measurement of exactly one ($\Pi_{Ai}$, $\Pi_{Bj}$) or two ($\Pi_{AiAj}$, $\Pi_{BiBj}$, $\Pi_{AiBj}$) photons, including any losses after the beamsplitter, are given by

$$\Pi_{Ai} = \eta_{Ai}c_i^\dagger|0\rangle\langle 0|c_i + \frac{1}{2}\eta_{Ai}(2-\eta_{Ai})c_i^{\dagger 2}|0\rangle\langle 0|c_i^2$$

$$+ \eta_{Ai}\sum_{j\neq i}(1-\eta_{Aj})(c_i^\dagger|0\rangle\langle 0|c_i) \otimes (c_j^\dagger|0\rangle\langle 0|c_j)$$

$$+ \eta_{Ai}\sum_j (1-\eta_{Bj})(c_i^\dagger|0\rangle\langle 0|c_i) \otimes (d_j^\dagger|0\rangle\langle 0|d_j), \quad (3a)$$

$$\Pi_{Bi} = \eta_{Bi}d_i^\dagger|0\rangle\langle 0|d_i + \frac{1}{2}\eta_{Bi}(2-\eta_{Bi})d_i^{\dagger 2}|0\rangle\langle 0|d_i^2$$

$$+ \eta_{Bi}\sum_{j\neq i}(1-\eta_{Bj})(d_i^\dagger|0\rangle\langle 0|d_i) \otimes (d_j^\dagger|0\rangle\langle 0|d_j)$$

$$+ \eta_{Bi}\sum_j (1-\eta_{Aj})(d_i^\dagger|0\rangle\langle 0|d_i) \otimes (c_j^\dagger|0\rangle\langle 0|c_j), \quad (3b)$$

$$\Pi_{AiAj} = \eta_{Ai}\eta_{Aj}(c_i^\dagger|0\rangle\langle 0|c_i) \otimes (c_j^\dagger|0\rangle\langle 0|c_j), \quad (3c)$$

$$\Pi_{BiBj} = \eta_{Bi}\eta_{Bj}(d_i^\dagger|0\rangle\langle 0|d_i) \otimes (d_j^\dagger|0\rangle\langle 0|d_j), \quad (3d)$$

$$\Pi_{AiBj} = \eta_{Ai}\eta_{Bj}(c_i^\dagger|0\rangle\langle 0|c_i) \otimes (d_j^\dagger|0\rangle\langle 0|d_j). \quad (3e)$$

We define unconditioned projectors

$$\Pi_{Ai}^{(s)} = \Pi_{Ai} + \sum_{j\neq i} \Pi_{AiAj} + \sum_{j} \Pi_{AiBj}$$

$$\approx \eta_{Ai} c_i^\dagger|0\rangle\langle 0|c_i + \eta_{Ai} \sum_{j\neq i} (c_i^\dagger|0\rangle\langle 0|c_i) \otimes (c_j^\dagger|0\rangle\langle 0|c_j)$$

$$+ \eta_{Ai} \sum_{j} (c_i^\dagger|0\rangle\langle 0|c_i) \otimes (d_j^\dagger|0\rangle\langle 0|d_j), \quad (4a)$$

$$\Pi_{Bi}^{(s)} = \Pi_{Bi} + \sum_{j\neq i} \Pi_{BiBj} + \sum_{i} \Pi_{AiBj}$$

$$\approx \eta_{Bi} d_i^\dagger|0\rangle\langle 0|d_i + \eta_{Bi} \sum_{j\neq i} (d_i^\dagger|0\rangle\langle 0|d_i) \otimes (d_j^\dagger|0\rangle\langle 0|d_j)$$

$$+ \eta_{Bi} \sum_{i} (c_i^\dagger|0\rangle\langle 0|c_i) \otimes (d_j^\dagger|0\rangle\langle 0|d_j), \quad (4b)$$

where we have neglected the possibility of two photons at identical frequencies. The ratio of losses $\eta_{Ai}$ and $\eta_{Bi}$ is then found to be equal to the ratio of the corresponding probabilities,

$$\frac{\eta_{Bi}}{\eta_{Ai}} = \frac{\text{Tr}(\rho\Pi_{Bi}^{(s)})}{\text{Tr}(\rho\Pi_{Ai}^{(s)})}. \quad (5)$$

**Single and coincidence rates**

Our amplitude and phase estimation relies on the assumption that the two photons have tight frequency correlations, so that $\Psi_{ij} = \psi_i \delta_{j,N+1-i}$. We also assume that the spontaneous parametric down-conversion (SPDC) emits a symmetric spectrum, $\psi_i = \psi_{N+1-i}$. The probabilities for an unconditioned single detection are then

$$\text{Tr}(\rho\Pi_{Ai}^{(s)}) = \frac{g}{2}\eta_{Ai}T_i|\psi_i|^2 + \frac{g}{2}\eta_{Ai}R_i|\psi_i|^2, \quad (6a)$$

$$\text{Tr}\left(\rho\Pi_{Bi}^{(s)}\right) = \frac{g}{2}\eta_{Bi}T_i|\psi_i|^2 + \frac{g}{2}\eta_{Bi}R_i|\psi_i|^2. \quad (6b)$$

The coincidence probabilities are

$$\text{Tr}\left(\rho\Pi_{AiA(N+1-i)}\right) = \frac{g}{4}\eta_{Ai}\eta_{A(N+1-i)}\left|\sqrt{T_iR_{N+1-i}}e^{i\phi_i} + \sqrt{T_{N+1-i}R_i}e^{i\phi_{N+1-i}}\right|^2|\psi_i|^2, \quad (7a)$$

$$\text{Tr}\left(\rho\Pi_{BiB(N+1-i)}\right) = \frac{g}{4}\eta_{Bi}\eta_{B(N+1-i)}\left|\sqrt{T_iR_{N+1-i}}e^{i\phi_i} + \sqrt{T_{N+1-i}R_i}e^{i\phi_{N+1-i}}\right|^2|\psi_i|^2, \quad (7b)$$

$$\text{Tr}\left(\rho\Pi_{AiB(N+1-i)}\right) = \frac{g}{4}\eta_{Ai}\eta_{B(N+1-i)}\left|\sqrt{T_iR_{N+1-i}}e^{i\phi_i} - \sqrt{T_{N+1-i}R_i}e^{i\phi_{N+1-i}}\right|^2|\psi_i|^2. \quad (7c)$$

Using the loss ratio $\eta_{Bi}/\eta_{Ai}$, we can define single and coincidence rates

$$S_i = \text{Tr}\left(\rho\Pi_{Ai}^{(s)}\right) + \frac{\eta_{Ai}}{\eta_{Bi}}\text{Tr}\left(\rho\Pi_{Bi}^{(s)}\right)$$

$$= g\eta_i(T_i + R_i)|\psi_i|^2, \quad (8a)$$

$$C_i^+ = \text{Tr}\left(\rho\Pi_{AiA(N+1-i)}\right) + \frac{\eta_{Ai}\eta_{A(N+1-i)}}{\eta_{Bi}\eta_{B(N+1-i)}}\text{Tr}\left(\rho\Pi_{BiB(N+1-i)}\right)$$

$$+ \frac{\eta_{A(N+1-i)}}{\eta_{B(N+1-i)}}\text{Tr}\left(\rho\Pi_{AiB(N+1-i)}\right) + \frac{\eta_{Ai}}{\eta_{Bi}}\text{Tr}\left(\rho\Pi_{AiB(N+1-i)}\right)$$

$$= g\eta_i\eta_{N+1-i}(T_iR_{N+1-i} + T_{N+1-i}R_i)|\psi_i|^2, \quad (8b)$$

$$C_i^- = \text{Tr}\left(\rho\Pi_{AiA(N+1-i)}\right) + \frac{\eta_{Ai}\eta_{A(N+1-i)}}{\eta_{Bi}\eta_{B(N+1-i)}}\text{Tr}\left(\rho\Pi_{BiB(N+1-i)}\right)$$

$$- \frac{\eta_{A(N+1-i)}}{\eta_{B(N+1-i)}}\text{Tr}\left(\rho\Pi_{AiB(N+1-i)}\right) - \frac{\eta_{Ai}}{\eta_{Bi}}\text{Tr}\left(\rho\Pi_{AiB(N+1-i)}\right)$$

$$= 2g\eta_i\eta_{N+1-i}\sqrt{T_iT_{N+1-i}R_iR_{N+1-i}}|\psi_i|^2\cos(\phi_i - \phi_{N+1-i}). \quad (8c)$$

These rates involve corrections for the relative efficiency $\eta_{Bi}/\eta_{Ai}$, and the final expressions use the shorthand $\eta_i = \eta_{Ai}$. Here, $S_i$ is the number of photons detected in frequency bin $i$, regardless of which fiber the photon emerged from; $C_i^+$ is the number of coincidence counts at frequency bins $i$ and $N + 1 - i$, also regardless of fiber; $C_i^-$ is the difference in coincidence counts between pairs emerging from the same fiber ("bunched pairs") and those emerging from opposite fibers ("antibunched pairs"), still at frequency bins $i$ and $N + 1 - i$. The rate $C_i^-$ therefore quantifies the phase-sensitive Hong-Ou-Mandel interference, while $S_i$ and $C_i^+$ can be used to estimate the absorption.

**Absorption estimation and calibration spectra**

Traditional quantum absorption measurements rely on the ratio of the singles rate and the coincidence rate, which is equal to the absolute transmission of the sample[1]. The expressions for $S_i$ and $C_i^+$ above show that the symmetrization imposed by the beamsplitter means that this ratio in the Hong-Ou-Mandel experiment mixes absorption at frequency bins $i$ and $N + 1 - i$, so that the coincidence-to-single ratio,

$$\frac{C_i^+}{S_{N+1-i}} = \eta_i \frac{T_i R_{N+1-i} + T_{N+1-i} R_i}{T_{N+1-i} + R_{N+1-i}}$$

is not simply related to the transmission at a single frequency. To separate the contributions from the two frequencies, it is necessary to perform calibration measurements without a sample present. For simplicity, we also assume that the sample of interest absorbs on only one half of the spectrum, which is a good approximation for the SiNC dye used here. Two calibration measurements are recorded, in addition to the measurements of the dye and reference samples. One calibration measurement places an opaque object in the sample arm; the other measurement places an opaque object in the reference arm.

We now discuss the estimator for the absorption. Below, we use $(s)$, $(r)$, $(0)$ and $(0')$ to denote expected rates in measurements with the dye sample ($T_i^{(s)} = e^{-\chi_i - \theta_i^{(s)}}$), reference sample ($T_i^{(r)} = e^{-\chi_i - \theta_i^{(r)}}$), blocked reference arm ($T_i^{(0)} = e^{-\chi_i}$), and blocked sample arm ($T_i^{(0')} = 0$), respectively. Here, $\chi_i$ denotes the losses due to scattering in the signal arm, and $\theta_i^{(s)}$ ($\theta_i^{(r)}$) describe the absorption due to the dye sample (reference sample). In all four measurements we assume that the transmission $R_i$ in the reference arm is unchanged, except in the case where it is blocked ($R_i^{(0)} = 0$); furthermore, we assume that the transmission $T_i$ through the dye and reference samples for frequency bins $i > N/2$ (corresponding to wavelengths longer than the degeneracy wavelength of 810 nm) is the same. We also assume that the pair probability $g$, the probe spectrum $\psi_i$, and the detection efficiencies $\eta_i$ remain unchanged between measurements (this assumption is discussed further in Supplementary Note 2). After performing all four measurements, we calculate the quantities

$$\frac{S_i^{(0)}}{S_i^{(0')}} = \frac{T_i^{(0)}}{R_i}, \quad (9a)$$

$$\frac{S_{N+1-i}^{(0)}}{S_{N+1-i}^{(0')}} = \frac{T_{N+1-i}}{R_{N+1-i}}, \quad (9b)$$

$$\frac{S_i^{(r)} - S_i^{(0')}}{S_i^{(0)}} = \frac{T_i^{(r)}}{T_i^{(0)}}. \quad (9c)$$

Using the fractional change in the coincidence-to-singles ratio,

$$\frac{C_i^{+(s)}/S_{N+1-i}^{(s)}}{C_i^{+(r)}/S_{N+1-i}^{(r)}} = \frac{T_i^{(s)}/T_i^{(r)} + T_{N+1-i}R_i/T_i^{(r)}R_{N+1-i}}{1 + T_{N+1-i}R_i/T_i^{(r)}R_{N+1-i}} \quad (10)$$

along with the correction term

$$\frac{S_{N+1-i}^{(0)}}{S_{N+1-i}^{(0')}} \frac{S_i^{(0')}}{S_i^{(0)}} \frac{S_i^{(0)}}{S_i^{(r)} - S_i^{(0')}} = \frac{T_{N+1-i}R_i}{R_{N+1-i}T_i^{(r)}}, \quad (11)$$

we can calculate the change in transmission between the reference and dye samples,

$$\frac{T_i^{(s)}}{T_i^{(r)}} = \left(1 + \frac{S_{N+1-i}^{(0)}}{S_{N+1-i}^{(0')}} \frac{S_i^{(0')}}{S_i^{(0)}} \frac{S_i^{(0)}}{S_i^{(r)} - S_i^{(0')}}\right) \frac{C_i^{+(s)}/S_{N+1-i}^{(s)}}{C_i^{+(r)}/S_{N+1-i}^{(r)}} - \frac{S_{N+1-i}^{(0)}}{S_{N+1-i}^{(0')}} \frac{S_i^{(0')}}{S_i^{(0)}} \frac{S_i^{(0)}}{S_i^{(r)} - S_i^{(0')}}. \quad (12)$$

This quantity, calculated using the experimentally measured rates, is the absorption estimate shown in black marks in Figure 3(a) of the main text.

**Phase estimation**

The phase of the probe beam, including the group delay of the cuvette and delay stage, is

$$\phi_i = \Phi_i - \omega_i(\tau_{DS} + \tau_C), \quad (13)$$

where $\Phi_i$ is the phase shift imparted by the sample, $\omega_i$ is the frequency of bin $i$, and $\tau_{DS}$ ($\tau_C$) are the delays imparted by the delay state (cuvette). We assume here that $\psi_i$, describing the spectrum of the down-converted light, is real, so that the phases resulting from the SPDC process itself may be neglected. For each frequency bin, the value $\Phi_i - \Phi_{N+1-i} - (\omega_i - \omega_{N+1-i})\tau_C$ is estimated by fitting the sinusoidal rate $C_i^-$, taken to be a function of $\tau_{DS}$.

At a fixed position of the delay stage, the relative phase has the form

$$\phi_i - \phi_{N+1-i} = \Phi_i - \Phi_{N+1-i} - (\omega_i - \omega_{N+1-i})(\tau_{DS} + \tau_C). \quad (14)$$

The cuvette delay $\tau_C$ takes different values for the dye sample's cuvette and the reference sample's cuvette. It is therefore necessary to correct for $\tau_C$ before comparing the measured phase shifts for the dye and reference samples. An estimate of this delay is obtained by calculating the sum $\sum_i C_i^-$ for each value of the stage delay $\tau_{DS}$. This sum is equivalent to the traditional "Hong-Ou-Mandel dip" seen in two photon interferometry experiments which are not frequency resolving[2]. If the sample does not impose a phase shift ($\Phi_i = \Phi_{N+1-i} = 0$), the expression for $C_i^-$ above shows that the sum is maximized when the cuvette delay is fully compensated, $\tau_{DS} = -\tau_C$. The phase shift $(\omega_i - \omega_{N+1-i})\tau_C$ is then subtracted from the fit values to obtain estimates of $\Phi_i - \Phi_{N+1-i}$. We find that even in the case of moderate phase shifts ($\Phi_i - \Phi_{N+1-i} \approx$ 0.5 rad), this procedure effectively corrects for the group delay of the cuvette. The differential phase between the dye ($s$) and reference ($r$) samples,

$$\left(\Phi_i^{(s)} - \Phi_{N+1-i}^{(s)}\right) - \left(\Phi_i^{(r)} - \Phi_{N+1-i}^{(r)}\right) \quad (15)$$

is calculated from the respective sets of measurements and a plot is shown in the main text. By assumption, the dye sample has significant absorption and phase shifts only on one half of the spectrum (wavelengths smaller than 810 nm), so that the differential phase may be entirely attributed to the phase shift $\Phi_i^{(s)} - \Phi_i^{(r)}$ imparted by the dye at a single frequency bin.

## 2. Systematic errors arising from fiber coupling

The current experiment relies on coupling photon pairs into single-mode fibers (SMF) before the spectrally resolving measurement. The spectrum of light produced by spontaneous parametric down-conversion is determined by energy and momentum conservation in the down-conversion process[3]. Energy conservation requires that the pump frequency $\omega_p$ is related to the down-converted frequencies $\omega_s$ and $\omega_i$ by $\omega_p = \omega_s + \omega_i$; this requirement has no explicit dependence on propagation direction. On the other hand, momentum conservation requires that the wavevectors mismatch $\delta\vec{k} = \vec{k}_p - \vec{k}_s - \vec{k}_i$ be small, where $\vec{k} = n(\hat{k})\omega/c$ and $\hat{k} = \vec{k}/|\vec{k}|$. In birefringent crystals, such as the PPKTP crystal used to produce photon pairs, the refractive index depends on the angle of propagation relative to the optic axes of the crystal. As a result, momentum conservation imposes a relationship between the pump and down-converted frequencies that changes with angle; a more detailed analysis shows that the intensity of down-converted light depends on the momentum mismatch via the function

$$\text{sinc}^2\left[\frac{L}{2}\left(\vec{k}_p - \vec{k}_s - \vec{k}_i - \vec{\Lambda}\right)_z\right], \qquad (16)$$

where $L$ is the crystal thickness and $\vec{\Lambda}$ is related to the crystal poling[4]. It follows that the angle-dependence in the momentum mismatch $\delta\vec{k}$ in turn causes an angle dependence to the down-converted intensity, so that the down-converted light generally has a correlation between propagation direction and spectrum.

In our experiment, light propagates through a glass cuvette containing the spectroscopic sample before coupling to single mode fiber. The effect of the single mode fiber is to reject any light with a propagation direction outside of a narrow range of values. This makes it easier to observe high Hong-Ou-Mandel visibility, but as a result of the spatial-spectral correlations also causes the measured spectrum to vary when the beam is deflected by a small amount. In placing cuvettes into the spectrometer, the cuvette angle is set so as to maximize the number of detected pairs (summed over all wavelengths) after the fiber; however, such a maximization does not guarantee that the spectrum of the coupled light is the same before and after inserting the cuvette, even in the absence of any frequency-dependent scattering or absorption by the sample.

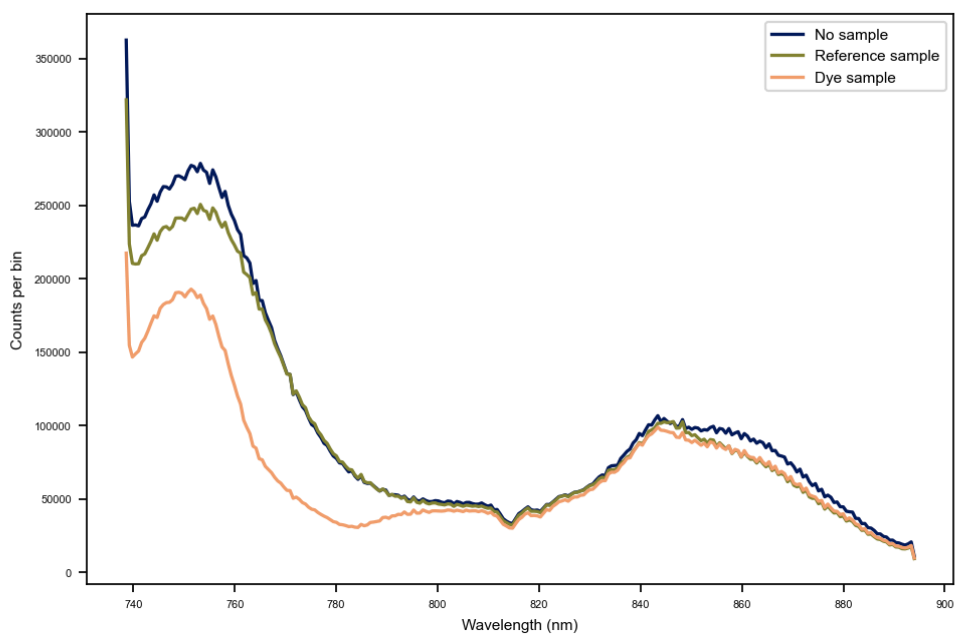

**Fig. S1**: Spectrum of unconditional detections. The values shown include both paired and unpaired detector clicks.

This change in spectrum is illustrated in Figure S1. Despite the reference sample (toluene) having negligible absorption at all measured wavelengths, there are significant differences between the measured spectrum for no cuvette and the spectrum for the reference sample. The effect of this coupling is even more apparent when looking at only paired detections (Figure S2); this is expected since coincidence detections occur only for tightly correlated spatial modes and so are generally more sensitive to misalignment than single detections.

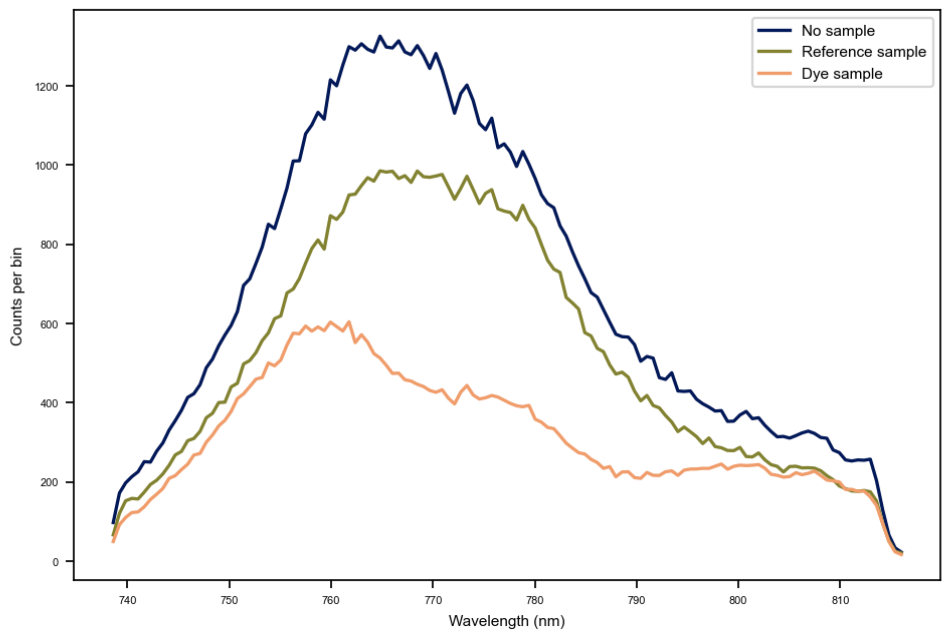

**Fig. S2**: Spectrum of coincidence detections. The values shown include only paired detector clicks.

In Figure 3(a) in the main text, we show the transmission curve obtained using two estimators. One relies on the ratio of coincidence counts to single counts, as described in Supplementary Note 1. The other estimator compares the number of single counts at each wavelength for measurements with and without a sample, as occurs in a traditional absorption spectrometer. The same set of measured data can then be used to obtain two separate absorption curves, with one curve relying on the coincidence and singles counts (the black marks in Figure 3(a)) and one curve using only singles counts (the red curve in the same figure). The two curves show similar systematic errors when compared to the reference (blue) curve, which supports the idea that these systematic errors are primarily due to optical misalignments introduced by the sample, rather than being a result of our choice of estimator. The beam deflection introduced by the cuvette appears primarily as a source of loss in the measured transmission spectrum, in addition to the expected loss which is due to absorption by the SiNC dye. A similar comparison cannot be made for the phase measurement, since phase information only appears in the coincidence spectra.

In any practical experiment involving fiber coupling of a parametric down-conversion source, the fiber will have some effect on the measured spectrum due to the spatial filtering it performs. Our spectrally-resolved Hong-Ou-Mandel (SRHOM) experiment is particularly vulnerable to these errors: the calculation of the SiNC dye's absorption (seen in the formula for $Q_i^{(s)}$ and $Q_i^{(r)}$ above) involves comparison between four separately measured spectra, two of which use unique cuvettes and two of which involve no cuvette. Hence, differences in the measured spectrum between calibration and sample measurements, and between the reference and dye samples, result in the systematic errors seen in the estimated absorption and phase profiles. Previous experiments demonstrating quantum-enhanced absorption spectroscopy use either multimode fibers[1] or do not use any fibers[5], which avoids this type of systematic error. Quantitative spectroscopy relying on SRHOM must address this issue, either by engineering a SPDC source that has uncorrelated spatial and spectral degrees of freedom, or by neglecting to use single mode fibers, which complicates the experimental setup but reduces this source of error.